**The outer hair cell of the organ of Corti possesses a voltage-dependent motile frequency response: evidence for the frequency-dependent compliance of prestin.**

**Joseph Santos-Sacchi [1,2,3]  Kuni Iwasa[4] and Winston Tan [1]**

*[1] Surgery (Otolaryngology), [2] Neuroscience, and [3] Cellular and Molecular Physiology, Yale University School of Medicine, 333 Cedar Street, New Haven, CT 06510, USA, [4] NIH-NIDCD*

***Running title:*** frequency-dependent effects of prestin's conformational compliance

***Send correspondence to:***
        Joseph Santos-Sacchi
        Surgery (Otolaryngology), Neuroscience, and Cellular and Molecular Physiology
        Yale University School of Medicine
        BML 224, 333 Cedar Street
        New Haven, CT  06510
        Phone: (203) 785-7566
        e-mail: joseph.santos-sacchi@yale.edu

**Acknowledgments**

This research was supported by NIH-NIDCD R01 DC000273, R01 DC016318 and R01 DC008130 to JSS.

**Abstract**

The outer hair cell (OHC) of the organ of Corti underlies a mechanically based process that enhances hearing, termed cochlear amplification. The cell possesses a unique motor protein, prestin, which senses voltage and consequently changes conformation to cause large cell length changes, termed electromotility (eM). In OHCs studied in vitro, the prestin voltage sensor generates a capacitance that is both voltage and frequency dependent, peaking in magnitude at a characteristic membrane voltage ($V_h$), which can be greater than the linear capacitance of the cell. Consequently, the OHC membrane time constant is multifarious depending upon resting potential and frequency of AC evaluation. After precisely correcting for this influence on the whole-cell voltage clamp time constant, we find that OHC eM is low pass in nature, substantially attenuating in magnitude within the frequency bandwidth of human speech. The frequency response is slowest at $V_h$, with a cut-off near 1.5 kHz, but increases up to six-fold in a U shaped manner as holding voltage deviates from $V_h$. NLC measures follow this pattern. Viscous drag alone cannot account for such eM behavior; nor can it arise from viscous drag in combination with a sigmoidal voltage-dependent OHC stiffness. However, viscous drag combined with kinetics of prestin, likely corresponding to its bell-shaped conformational gating compliance (Iwasa, 2000), is in line with our observations. How OHC eM influences cochlear amplification at higher frequencies needs reconsideration.

Outer hair cell (OHC) electromotility (eM) underlies cochlear amplification in mammals, where in its absence hearing deficits amount to 40-60 dB (Dallos et al., 2008; Ashmore et al., 2010). The molecular basis of OHC eM is the membrane bound protein prestin (SLC26a5), an anion transporter family member that has evolved to work as a voltage-dependent motor protein in these cells (Zheng et al., 2000). The protein's voltage-sensor activity presents as a voltage dependent (or nonlinear) capacitance (NLC), obeying Boltzmann statistics (Ashmore, 1990; Santos-Sacchi, 1991), and whose peak magnitude corresponds to the voltage ($V_h$) where sensor charge is equally displaced to either side of the membrane. Voltage-dependent conformational change in prestin is believed to form the basis of precise phase differences between OHC activity and basilar membrane motion that leads to amplification, indicative of local cycle-by-cycle feedback (Dallos et al., 2008). However, recent evidence showing an apparent absence of such phase differences has challenged this view (Ren et al.). Additionally, it is commonly accepted that unconstrained (absent load) OHC eM magnitude is invariant across stimulating frequency, having been measured out beyond 80 kHz (Frank et al., 1999); the flat frequency response is another important element in cycle-by-cycle feedback theory. However, these data were obtained at voltage offsets far removed from $V_h$, which has questionable physiological significance. By exploring across a range of holding potentials, passing through $V_h$,  it was recently demonstrated that eM displays low-pass behavior (Santos-Sacchi and Tan, 2018). Here we further explore the low-pass nature of eM under whole cell voltage clamp out to 6.25 kHz with ramped changes in holding potential that provide shorter, more stationary-in-time measures compared to our previous experiments. We find that eM and NLC show very good agreement in their nonlinear voltage and frequency dependence, both frequency cut-offs being U-shaped functions of holding voltage, with minima at $V_h$. Our data point to a viscoelastic relaxation process, involving both viscous drag on the OHC and mechanical compliance of the OHC, that determines the frequency dependence of eM in our configuration. This behavior arises because "gating compliance" associated with prestin's conformational transitions must have the same voltage and frequency dependence as NLC (Iwasa, 2000).

**Methods**

Whole-cell recordings were made from single isolated OHCs from the apical two turns of organ of Corti of guinea pigs. An inverted Nikon Eclipse TI-2000 microscope with 40X lens was used to observe cells during voltage clamp. Experiments were performed at room temperature. Blocking solutions were used to remove ionic currents, limiting confounding effects on NLC determination and voltage delivery under voltage clamp (Santos-Sacchi, 1991; Santos-Sacchi and Song, 2016). Extracellular solution was (in mM): NaCl 100, TEA-Cl 20, CsCl 20, CoCl$_2$ 2, MgCl$_2$ 1, CaCl$_2$ 1, Hepes 10. Intracellular solution was (in mM): CsCl 140, MgCl$_2$ 2, Hepes 10 and EGTA 10. All chemicals were purchased from Sigma-Aldrich.

An Axon 200B amplifier was used for whole-cell recording with jClamp software (www.scisoftco.com). An Axon Digidata 1440 was used for digitizing. AC analysis of membrane currents (I$_m$) and eM were made by stimulating cells with a voltage ramp from 100 to -110 mV (nominal), superimposed with summed AC voltages at harmonic frequencies from 195.3 to 6250 Hz, with a 10 μs sample clock. Currents were filtered with a 4-pole Bessel filter. Corrections for series resistance were made during analysis. Magnitude and phase of responses were computed by FFT in Matlab. Currents were corrected for the frequency response of the voltage clamp system. Capacitance was measured using dual-sine analysis at harmonic frequencies (Santos-Sacchi et al., 1998; Santos-Sacchi, 2004). In order to extract Boltzmann parameters, capacitance-voltage data were fit to the first derivative of a two-state Boltzmann function.

$$C_m = Q_{\max} \frac{ze}{k_B T} \frac{b}{(1+b)^2} + C_{\lin} \quad \text{where} \quad b = exp\left(-ze\frac{V_m - V_h}{k_B T}\right) \qquad \text{(m.1)}$$

Q$_{max}$ is the maximum nonlinear charge moved, V$_h$ is voltage at peak capacitance or equivalently, at half-maximum charge transfer, V$_m$ is R$_s$-corrected membrane potential, $z$ is valence, C$_{lin}$ is linear membrane capacitance, e is electron charge, $k_B$ is Boltzmann's constant, and T is absolute temperature.

NLC determination was made following stray capacitance removal (Santos-Sacchi et al., 1998; Santos-Sacchi and Song, 2016; Santos-Sacchi and Tan, 2018). Patch pipettes were coated with M-coat (Micro Measurements, PA) to reduce stray capacitance. Remaining stray capacitance was removed by amplifier compensation circuitry prior to whole-cell configuration, and if necessary additional compensation was applied under whole cell conditions (Schnee et al., 2011a, b) and/or through a software algorithm to ensure expected frequency-independent linear capacitance (Santos-Sacchi, 2018). Series resistance ($R_s$) was determined from voltage step-induced whole cell currents, the derivation provided in the Appendix of (Huang and Santos-Sacchi, 1993). $R_m(v)$ across ramp voltage was determined from $R_s$-corrected ramp voltage and generated ramp currents with AC components removed, $\Delta V_m/(\Delta I_{Rs})$.

Simultaneous and synchronous eM measures were made with fast video recording. A Phantom 110 camera (Vision Research) was used for video measures at a frame rate of 25 kHz. Magnification was set to provide 176 nm/pixel. A method was developed to track the apical image of the OHC, providing sub-pixel resolution of movements (see **Fig. 1**). Video frames were filtered with a Gaussian Blur filter ([www.gimp.org](www.gimp.org)) prior to measurement. The patch electrode provided a fixed point at the basal end of the cell. Resultant movements were analyzed by FFT in Matlab. The colors in surface plots were generated in the Matlab plotting routine *surf* with shading set to <interp>. This procedure allows contours to be readily observable. Although eM was measured at all 6 AC frequencies, only 5 estimates of NLC were possible using the dual-sine approach. In order to obtain $\tau_{clamp}$ estimates at 6.25 kHz, the magnitude of NLC at 3.125 kHz was used, factored by 0.8.

Models were implemented in Matlab Simulink and Simscape, as detailed previously (Song and Santos-Sacchi, 2013; Santos-Sacchi and Song, 2014a). Model $R_s$ was 7.12 M$\Omega$. $R_m$ was 200 M$\Omega$, $C_{lin}$ was 23 pF. $R_s$-corrected $V_m$ vs. eM data were fit with the first derivative of a 2-state Boltzmann function (Santos-Sacchi, 1991). Data are presented as mean +/- se.

Under voltage clamp, the voltage delivered across the cell membrane depends on the electrode series resistance ($R_s$). In the absence of $R_s$, the command voltage is delivered faithfully to the cell membrane in magnitude and time (phase). Otherwise, the voltage delivered to the membrane will suffer from voltage drops across $R_s$, depending on the magnitude and time (phase) of evoked currents. In any analysis of voltage-dependent cellular processes, such as electromotility (Santos-Sacchi and Dilger, 1988; Iwasa and Kachar, 1989), it is important to accurately assess membrane voltage. Actual membrane potential under voltage clamp can be exactly determined by subtraction of the voltage drop across $R_s$, i.e., $I_{Rs} * R_s$, with $I_{Rs}$ being the sum of resistive and capacitive components of the cell membrane current. For sinusoidal stimulation across the voltage ramp, the AC command voltage $V_c$ and evoked currents $I_{Rs}$, are evaluated as complex values at each excitation frequency and ramp offset voltage, $(A + jB)$, where A and B are the real and imaginary components, obtained by FFT, thereby supplying $V_c(f, v)$ and $I_{Rs}(f, v)$. Our goal is to accurately determine the frequency response of eM, given a non-zero series resistance that imposes its own influence on the frequency response of voltage-driven eM data. The method essentially seeks the true excitation voltage $V_m(f, v)$, i.e., the drive for eM, supplied to the plasma membrane.

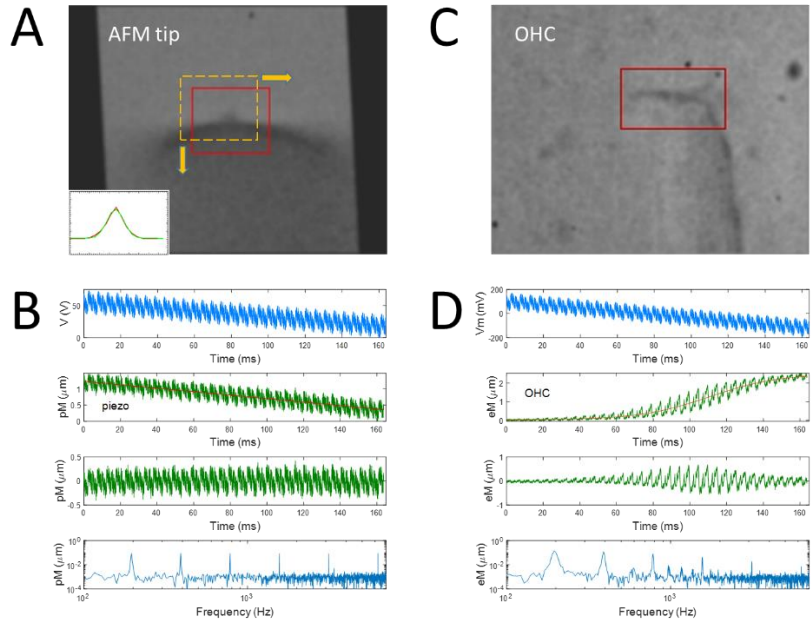

**Fig. 1** Motion measures from fast video using a shape tracking algorithm implemented in the program jClamp. **A)** Frame of an AFM probe tip with a fiduciary red box delimiting a region of the image to track. For each frame in the video stream, the image was sampled at single pixel offsets of +/- 10 pixels in both the vertical and horizontal dimensions (yellow dotted box) relative to the fiducial image. Rotation of the image was possible to maximize displacements to a single dimension. A pixel comparison was made (subtraction of sampled image pixels from fiducial pixels and summed). The resulting distribution along the x and y axis was fit with a Gaussian function to obtain subpixel image movements. **B)** *Panel 1* shows the voltage command to the piezo stack. *Panel 2* shows the piezo movements (pM) induced by the voltage command. The red line is a linear fit used to detrend the data (*Panel 3*), prior to FFT of the whole duration response (*Panel 4*). The magnitude plot shows high fidelity in capturing tip movements across frequency. **C)** The apical portion of an OHC under whole cell voltage clamp was similarly tracked. **D)** *Panel 1* shows voltage clamp command. *Panel 2* shows OHC eM. The red line is a sigmoidal fit used to detrend the data (*Panel 3*), prior to FFT of the whole duration response (*Panel 4*). Unlike the piezo response, eM falls precipitously across frequency. The frequency response arises from $\tau_{clamp}$ and the intrinsic low pass nature of eM. The removal of $\tau_{clamp}$ contribution is easily accomplished (see **Results**).

Before we present an analysis of averaged eM data, we illustrate the approaches available for $R_s$ correction in a Matlab Simulink model. In the model, we are able to directly measure imposed membrane voltage so we can compare results from these approaches to actual values. The generated charge in the model is taken as the equivalent of eM, since we have shown that OHC eM and prestin charge movement are directly coupled (Santos-Sacchi and Tan, 2018). The model we use has been fully described before (Santos-Sacchi and Song, 2014b), and includes a slow, stretched exponential intermediary transition between chloride binding transitions and voltage-dependent charge transitions, the latter corresponding to eM. Modifications of the model were made to correspond to the biophysical data. Chloride was 140 mM. The number of prestin particles was set to 18e6, and $z = 1$. The forward and backward voltage-dependent transition rate constants were 1.2947e6 and 1.1558e4, respectively. Both forward and backward parallel intermediary transition rate constants were defined as

$$\alpha, \beta = A \cdot exp(b \cdot m), form = 0..26. \qquad \text{(m.2)}$$

$$\text{where } b = \text{-0.4663 and } A = 3.0398e4.$$

For the model, there are three ways to correct the frequency response of eM magnitude for the confounding effects of $R_s$.

***Method 1***) The complex ratio of AC command voltage (located at differing ramp voltage offsets [$v$]) to the *directly* measured membrane voltage within the model at each excitation frequency can be used to correct eM magnitude frequency response, eM being both a function of frequency and holding voltage (Santos-Sacchi and Tan, 2018), namely, eM $(f, v)$, akin to NLC $(f, v)$ (Santos-Sacchi and Song, 2016). Parallel bars indicate absolute values.

$$|eM(f, v)_{actual}| = \left| eM(f, v)_{measured} \cdot \frac{V_c(f, v)}{V_m(f, v)} \right| \qquad \text{(m.3)}$$

***Method 2***) The complex ratio of command voltage to calculated $V_m(f, v)$, namely $(I_{Rs}(f, v) \cdot R_s$, can be used for correction,

$$|eM(f, v)_{actual}| = \left| eM(f, v)_{measured} \cdot \frac{V_c(f, v)}{(V_c(f, v) - I_{Rs}(f, v) \cdot R_s)} \right| \qquad \text{(m.4)}$$

*Method 3*) Finally, the multifarious clamp time constant, $\tau_{clamp}(f, v)$, determined at each excitation frequency and ramp voltage offset can be used to correct, via a Lorentzian function ($A$ = $1/[1 + (2\pi f\tau)^2]^{1/2}$), the magnitude of eM, scaled to DC levels $[1 - R_s/R_m(f_0, v)]$, $f_0 = 0 \approx$ ramp frequency.

$$\tau_{clamp}(f, v) = [C_{lin} + NLC(f, v)] \cdot R_{||}(f_0, v) \qquad \text{(m.5)}$$

$$\text{where } R_{||}(f_0, v) = \frac{R_s * R_m(f_0, v)}{R_s + R_m(f_0, v)}$$

$$|eM(f, v)_{actual}| = |eM(f, v)_{measured}| \cdot \frac{[1 + (2\pi f \cdot \tau_{clamp}(f, v)^2]^{1/2}}{[1 - \frac{R_s}{R_m(f_0, v)}]} \qquad \text{(m.6)}$$

For the model, Supplemental **Fig. 1** shows that each method of $R_s$ correction gives precisely the same results, revealing the intrinsic low pass eM response of the meno presto model that derives from its transition kinetics (Song and Santos-Sacchi, 2013; Santos-Sacchi and Tan, 2018).

## Results

**Fig. 1** depicts the method used to measure eM. Movement of a piezo-driven AFM tip (0.2 N/m) confirmed the fidelity of the measuring technique out to 6250 Hz (**Fig. 1 B**, *Panel 4*). In order to obtain accurate FFT results, the ramp-induced movement was detrended by subtracting a linear fit for the AFM tip measures or a sigmoidal fit for the OHC eM measures (**Fig. 1 B** *Panel 3*, **D** *Panel 3*). Subsequent analyses followed this detrending approach. In **Fig. 1D** *Panel 4*, an FFT of the whole ramped eM response was made, and shows that unlike the piezo-driven response (**Fig. 1B** *Panel 4*), the magnitude of OHC eM falls precipitously with frequency. This roll-off arises from both the voltage-filtering effects of $R_s$ and the kinetics of prestin. In the following analysis, we restrict FFT analysis to defined integral segments (see **Supplemental Figure** and **Fig. 3;** red and blue highlighted example regions) of the eM ramp response to assess voltage dependence. Furthermore, we detail exact methods (see **Methods**) to remove the effects of series resistance interference, thus revealing true eM frequency response.

For the OHC under voltage clamp we do not have direct access to the membrane voltage as in the model (see Methods), so only two methods for eM correction are available – that using estimates of $R_s$ and NLC to gauge $\tau_{clamp}$, and that using direct measures of whole cell currents. **Fig. 2** shows $R_s$ corrected and uncorrected AC command voltage delivered to the cells. Correction precisely accounts for the frequency-dependent roll-off in AC voltages, the same correction being applied to voltage-dependent eM measures.

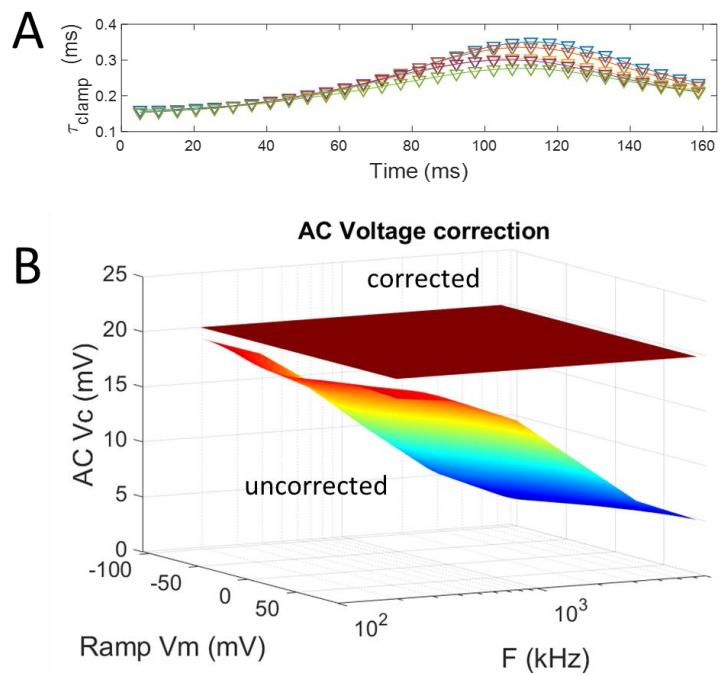

**Fig. 2** **A)** A multifarious clamp time constant arises from the frequency and voltage dependent nature of OHC nonlinear capacitance ($\tau_{clamp}$ = $R_s$ * NLC). **B)** Complex whole cell currents or calculated multifarious clamp time constants were used to correct for the roll-off in AC command voltage (see **Methods**). Each approach gives equivalent results.

We chose 4 cells for analysis with very similar electrical characteristics and with eM commensurate with previously reported eM gains of about 15 nm/mV (Ashmore, 1987; Santos-Sacchi, 1989). Average (+/- se) Boltzmann parameters of NLC at the lowest frequency of 195 Hz are $Q_{max}$: 3.12 +/- 0.07 pC; $V_h$ -45.5 +/- 1.9 mV; z: 0.90 +/- 0.03; and $C_{lin}$ 23.22 +/- 0.6 pF. $R_s$ was 7.25 +/- 0.25 MΩ. Average eM gain evoked by the ramp protocol (**Fig. 3 A** *middle panel*) was 14.3 +/- 3.0 nm/mV. **Fig. 3** surface plots show average OHC eM before (**a**) and after (**b, c**) correcting for $R_s$-induced voltage errors. Utilizing either of the two possible approaches produces quite similar results confirming the low pass nature of eM. Notably, the equivalence of the two methods also confirms the validity of the dual sine approach to measure high frequency capacitance, whose multifarious influence on clamp time constant must be considered in order to properly evaluate eM frequency response under voltage clamp, and must be considered an issue for receptor potential generation *in vivo*, as well.

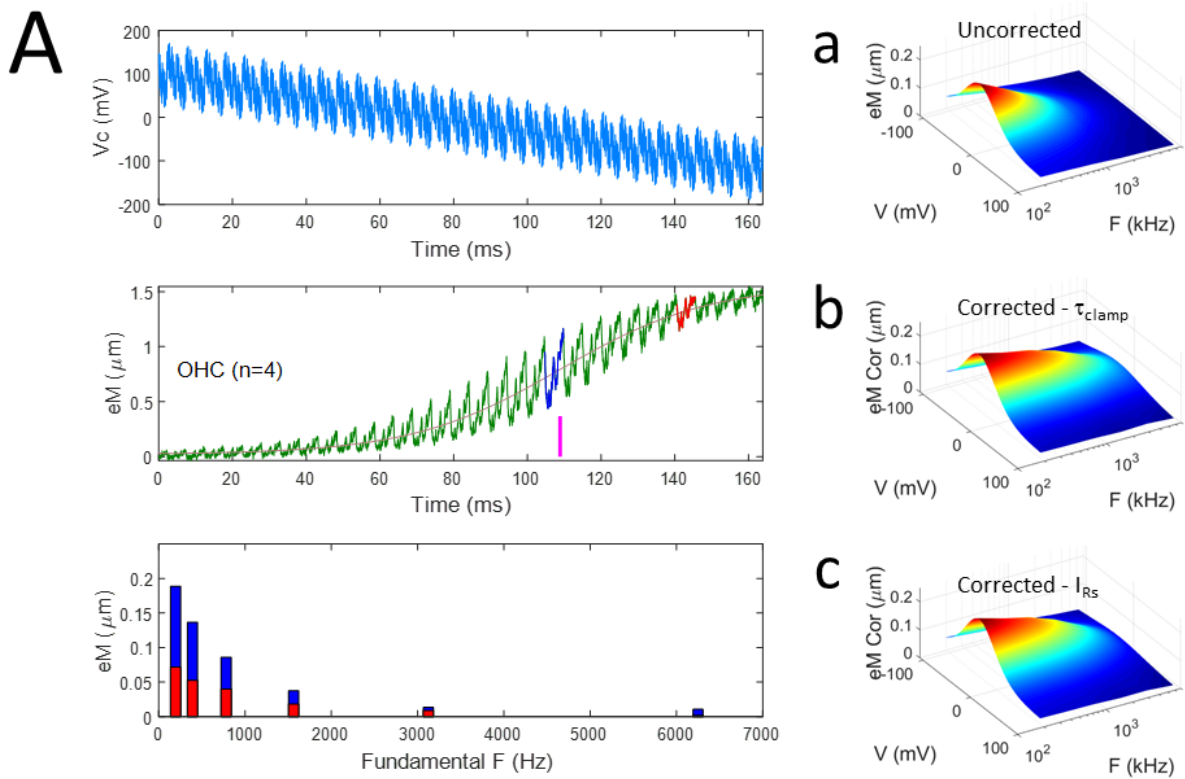

**Fig. 3** Average OHC eM frequency response corrected by 2 methods. **A)** *Panel 1* shows the hyperpolarizing ramp voltage stimulus (100 to -120 mV) with superimposed sum of 20 mV AC harmonic frequencies (195.3 – 6250 Hz). A pedestal of 100 mV (81.32 ms) preceded the ramp. *Panel 2* shows raw eM. Two selected integral regions are highlighted. Blue is near $V_h$ (magenta bar beneath shows position of $V_h$). Red highlights a hyperpolarized region. *Panel 3* shows magnitude responses at the 6 interrogated frequencies. Data were detrended as in **Fig. 1** before FFT. The surface plots on the right provide a full analysis across ramp voltage at all integral regions of the eM data. **a)** Uncorrected eM with average $R_s$: 7.12 MΩ; **b)** eM adjusted with *Method 3* based on corrections to multifarious $\tau_{clamp}$ (see Results). **c)** eM adjusted with *Method 2* based on $V_m$ calculation, namely, $V_m = V_c - I_{Rs} * R_s$. Note correspondence between the two approaches. OHC eM is low pass in nature after corrections.

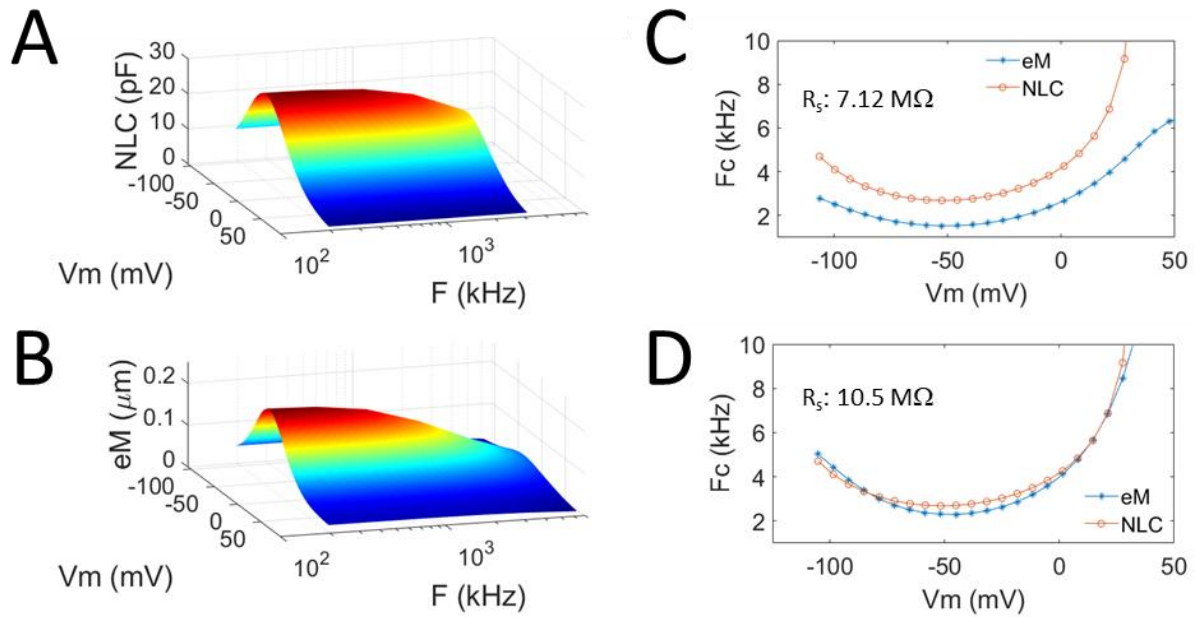

**Fig. 4 A, B)** Single Lorentzian fit of the eM data in **Fig. 3c** and its associated NLC. The surface plots show that the frequency cut-off ($F_c$) of eM and NLC increases as ramp voltage deviates from $V_h$. The cutoff is a U shaped function of holding voltage, eM being about 1.5 kHz and NLC being about 2.7 kHz at $V_h$ **(C)**. This slight mismatch could be accounted for by drag on the chamber bottom. However, if the $R_s$ estimate is increased from 7.12 M$\Omega$ to 10.5 M$\Omega$ then the correspondence is quite good **(D)**. $R_s$ is commonly found to increase during the course of an experiment.

In **Fig. 4A and B,** we show single Lorentzian fits of the eM data in **Fig. 3C** and its associated NLC. The plots illustrate the changing frequency response as ramp voltage passes through $V_h$. The cut-off frequencies are plotted in **Fig. 4C**, and displays U-shaped dependencies on holding potential, AC voltage stimulation at $V_h$ providing the slowest response. This is not expected from viscosity effects on cell movements, as such effects should be voltage independent, but interestingly corresponds to the expected compliance of the OHC (Iwasa, 2000). The disparity in the frequency responses is small, but if we assume that $R_s$ has increased a few M$\Omega$ during the course of the experiment, as commonly happens, the disparity disappears (**Fig. 4D**). Thus, the kinetics of NLC (likely corresponding to prestin compliance (Iwasa, 2000)) account for the voltage-dependence of eM frequency response.

## Discussion

Prestin works by sensing voltage with consequent alterations in its conformational state, leading to contractions and elongations of the cylindrical cell that provides enhancement of auditory threshold (Ashmore, 2008; Santos-Sacchi et al., 2017). Because of the large restricted prestin-generated charge movements evoked by voltage (Ashmore, 1990; Santos-Sacchi, 1991), a substantial alteration in membrane capacitance ensues, paradoxically altering the very voltage that the sensor senses. This is even expected to occur during normal acoustic stimulation, where the receptor potential will suffer from a multifarious membrane RC filter. For example, the receptor potential cutoff frequencies estimated using linear capacitance alone (Johnson et al., 2011), would be greatly reduced at $V_h$, where NLC can be as great as linear capacitance. In our study under voltage clamp, series resistance and membrane capacitance conspire to limit the imposition of voltage across the membrane, interfering in both time and magnitude. In order to evaluate effects of membrane potential across frequency it is required to alleviate this interference. Here we have compensated for the multifarious time constants induced by $R_s$ and $NLC(v,f)$ in order to reveal the true low-pass frequency dependence of OHC eM, evident within the frequency bandwidth of human speech. Thus, the OHC eM frequency response roll-off is not due to the frequency response of the driving voltage, but results from other factors intrinsic and extrinsic to the cell. As we argue below these factors are the viscosity of the medium bathing the cell and prestin's intrinsic compliance, corresponding to the protein's transition kinetics.

Length changes of the OHC in fluid are subject to viscous drag. The amplitude depends on the stiffness of the material that produces the motile force. In calibration experiments, force is produced by a stiff piezoelectric actuator (**Fig. 1A**), and no roll-off was observed. In our biophysical experiments, force was produced by a more compliant OHC (**Fig.1B, Fig. 3**), resulting in the observed frequency roll-off. This indicates that viscoelastic relaxation produced a low pass filter. In the following, a quantitative discussion is given.

Viscoelastic relaxation gives rise to a cut-off frequency, $f_c$, which is determined by $2\pi f_c = k/\eta$, where $k$ is the stiffness of the cell and $\eta$ is drag coefficient. First, let us examine the magnitude of the effect. For 60 µm long OHCs, experimental values for the axial stiffness range between 1.3 mN/m and 200 mN/m with a mean value of 4 mN/m (Hallworth, 1995). Using more stringent OHC selection criteria, a less variable value of 510±100 nN/m per unit strain (Iwasa and Adachi, 1997) gives rise to the stiffness of 8.5±1.7 mN/m. The drag on the cell would be between the Stokes drag on a sphere of 5µm radius and that of 10µm radius, given that the radius of the cylindrical cell body is about 5 µm. If we use the stiffness of 8.5 mN/m for the drag on a 5µm sphere, the cut-off frequency is 16 kHz, and 8 kHz for a 10 µm sphere. With the stiffness of 4 mN/m, the cut-off frequency is 8 kHz for a 5 µm radius sphere and 4 kHz for 10 µm radius. **Fig. 4C and D** shows that the cut-off frequency at $V_h$ is about 1.5 - 4 kHz, in a range not incompatible with the above calculations, particularly so after including "gating compliance" as discussed below.

Next, we may assume that the drag coefficient $\eta$ is constant. Then the roll-off frequency is related to the stiffness $k$ or compliance $1/k$ of the cell. The U-shaped voltage dependence of the cut-off frequency implies that the stiffness must have a U-shaped frequency dependence. In other words, the compliance $1/k$ is bell-shaped, similar to nonlinear capacitance. Indeed, **Fig. 4 (C, D)** shows the voltage dependence of the roll-off frequency and that of nonlinear capacitance are in agreement.

Such agreement is associated with the piezoelectric nature of eM (Iwasa, 2000). $L$ and $Q$ are the length and charge of an OHC, respectively. In a two-state conformational formulation (i.e., expanded-contracted motile element), these quantities can be expressed as functions of the force, $F$, applied to the cell and its holding membrane potential, $V$, in the following way:

$$L(F,V) = L_0 + gF + aNP(F,V), \qquad (1)$$

$$Q(F,V) = Q_0 + C_{lin}V + qNP(F,V), \qquad (2)$$

where $a$ and $q$ ($= ze$) are, respectively, the mechanical and the electrical displacement due to the unit motile element. The total number of motile elements is $N$. The quantity $P(F,V)$ is the fraction of the motile element in the expanded state and can be expressed by a Boltzmann function

$$P(F,V) = \frac{exp[-(q(V - V_h) + aF)/k_B T]}{1 + exp[-(q(V - V_h) + aF)/k_B T]}. \qquad (3)$$

Notice that the inclusion of a mechanical energy term, *aF*, introduces piezoelectric characteristics to the OHC. These equations lead to the compliance *1/k* (=*dL/dF*) of the cell and the membrane capacitance *dQ/dV*:

$$dL/dF = g + a^2 NB(F,V)/k_B T, \qquad (4)$$

$$dQ/dV = C_{lin} + q^2 NB(F,V)/k_B T, \qquad (5)$$

with

$$B(F,V) = \frac{exp[-(q(V - V_h) + aF)/k_B T]}{(1 + exp[-(q(V - V_h) + aF)/k_B T])^2}. \qquad (6)$$

The first term in *Eq. 4* is the material compliance and the second term an analogue of "gating compliance" (Howard and Hudspeth, 1987). In *Eq. 5*, the first term is the standard linear membrane capacitance and the second term is NLC. Notice that gating compliance and NLC have exactly the same functional dependence.

The above formulas are for the quasi-static case. The frequency dependence of NLC appears as a multiplying factor $f_c^2 / (f^2 + f_c^2)$ to the function *B(F,V)*. Here $f_c = k/(2\pi\eta)$ is the characteristic frequency of viscoelastic relaxation, where *η* is the drag coefficient. This expression is obtained as a special case (absence of both external elastic and inertial load) from a more general equation (Iwasa, 2016). The symmetry of the Boltzmann function with respect to electrical and mechanical terms applies the same factor to the compliance 1/k.

For these reasons, we may interpret that the cut-off frequency that we observed is determined by a viscoelastic effect involving isolated OHCs, viz., the interaction of viscous drag and cell compliance resulting from prestin activity. Our analysis further suggests that the study of OHC eM additionally requires considering the reciprocal effect arising from mechanical factors, in common with studies on other piezoelectric materials.

OHC eM has been taken to have a flat frequency response out to 80 kHz (Frank et al., 1999). However, those observations of free movements in the microchamber were made at voltage offsets far removed from $V_h$, which has questionable physiological significance. Additionally, force measures in those experiments were obtained under virtually isometric conditions, thus not providing frequency estimates that would result from a more physiological state where the OHC can transfer energy to its surrounding tissue. Though free movement of the isolated OHC also suffers from this limitation, here we use these measures to identify factors that control the frequency response, providing insights into the cell's behavior under more physiological conditions.

We recently showed that the frequency response of the OHC using the microchamber approach actually depends on the degree of voltage offset from $V_h$ (Santos-Sacchi and Tan, 2018). In that prior study, we compensated for clamp time constant induced errors using a single clamp time constant determined at each offset voltage, ignoring possible frequency-dependent effects on clamp time constant. With appropriate corrections, these observations are borne out in our present study. Thus, the relationship between OHC resting potential and NLC $V_h$ will govern the frequency response of eM, as well as the frequency response of voltage drive to prestin during acoustic generation of receptor potentials *in vivo*. We previously proposed that a mismatch between resting potential and NLC $V_h$ could enhance the frequency response of eM through a gain-bandwidth adjustment. Our new data extend and confirm this proposal. Furthermore, we detail the resting potential dependence of eM frequency response cut-off, being U shaped about $V_h$, and changing about six fold as holding voltage deviates from $V_h$. The voltage dependence is mirrored by the frequency roll-off of NLC. This correspondence is consistent, as we have argued, with the piezoelectric nature of OHC motility in which NLC is the electric counterpart of prestin's gating compliance (Iwasa, 2000). Therefore, our observation further validates the piezoelectric nature of OHC eM in the frequency domain.

Interestingly, our observation is at odds with the notion of a voltage dependent stiffness that mirrors the shape of a sigmoidal Boltzmann function (He and Dallos, 1999, 2000). Hallworth (Hallworth, 2007) did not find evidence for such voltage-dependent stiffness in OHCs. The U shaped frequency response, however, is compatible with a compliance of similar shape that arises from prestin. Our observations also demonstrate that the measured low-pass eM is not simply an intrinsic property of prestin and its influence on the cell's mechanical compliance, but the result of an additional viscous load, forming a viscoelastic filter. Therefore, the physiological frequency limit of OHC may be determined only by imposing a mechanical load, analogous to *in vivo* condition. Whether a voltage-dependent frequency response might assist cochlear amplification requires new cochlear models that incorporate our observations. Nevertheless, the influence of OHC activity on cochlear amplification is more complicated than has been envisioned.

Finally, we note that the kinetics of prestin are readily monitored through shifts in $V_h$, a reflection of the ratio of forward to backward transition rates. These kinetics, a determining factor in eM frequency response, depend on a host of other factors, including intracellular chloride, membrane tension and thickness, and temperature (Iwasa, 1993; Gale and Ashmore, 1994; Meltzer and Santos-Sacchi, 2001; Oliver et al., 2001; Santos-Sacchi et al., 2001; Izumi et al., 2011; Santos-Sacchi and Song, 2016). Indeed, alterations in prestin kinetics have been found in mutations in prestin (Homma et al., 2013). All these data implicate corresponding alterations in prestin's gating compliance, as well. We suspect that eM frequency response is not a static feature *in vivo*.

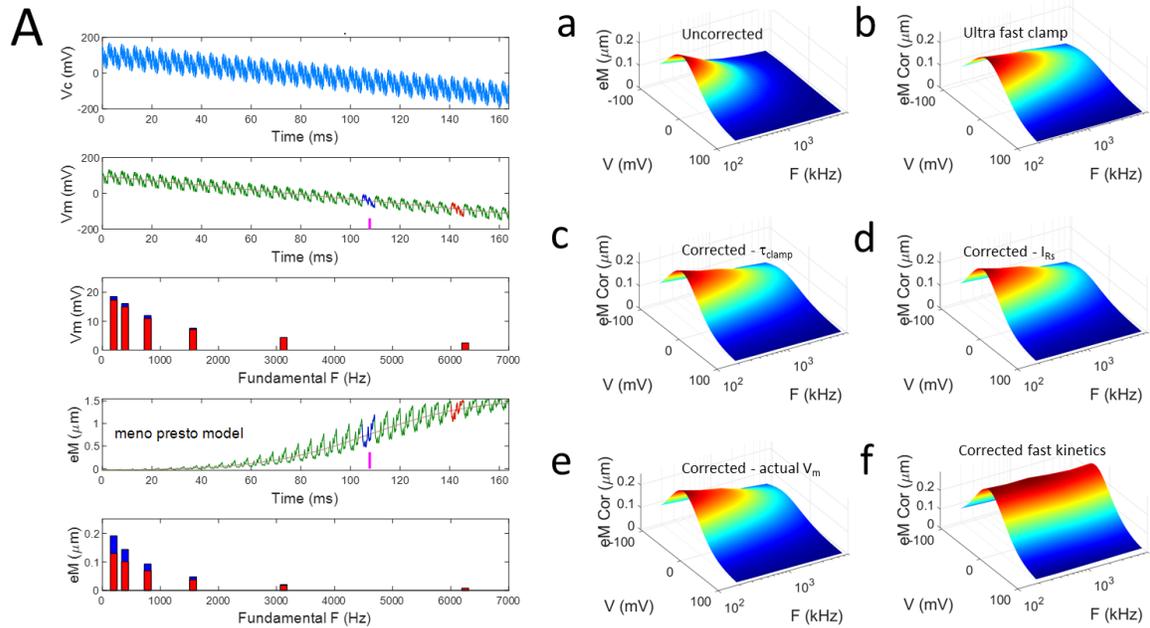

**Supplemental Figure** *Meno presto* model of OHC under voltage clamp illustrating 3 methods for eM correction. **A)** *Panel 1* shows the hyperpolarizing ramp voltage stimulus (100 to -120 mV) with superimposed sum of 20 mV AC harmonic frequencies (195.3 – 6250 Hz). A pedestal of 100 mV (81.32 ms) preceded the ramp. *Panel 2* shows membrane voltage. Two selected integral regions are highlighted. Blue is near $V_h$ (magenta bar beneath shows position of $V_h$). Red highlights a hyperpolarized region. In *Panel 3*, these regions were analyzed by FFT and magnitude results shown. *Panel 4* shows raw eM (scaled charge from model) also depicting selected regions to analyze. *Panel 5* shows magnitude responses at the 6 interrogated frequencies. Data were detrended as in **Fig. 1** before FFT.

The surface plots on the right provide a full analysis across ramp voltage at all integral regions of the eM data. eM-$V_m$ data at each frequency were fit with the first derivative of a 2-state Boltzmann prior to plotting. **a)** Uncorrected eM with $R_s$: 7.12 M$\Omega$; **b)** Uncorrected eM with a fast $\tau_{clamp}$ due to $R_s$: 0.01 M$\Omega$ (slowest $\tau_{clamp}$ at $V_h$ of 1.2 $\mu$s or $F_c$ of 133 kHz). Note frequency response of eM is faster than under the higher $R_s$ condition; **c)** eM (with $R_s$: 7.12 M$\Omega$) adjusted with **Method 3** based on corrections to multifarious $\tau_{clamp}$ (see Results). Note similarity of frequency response to fast $\tau_{clamp}$ results; **d)** eM (with $R_s$: 7.12 M$\Omega$) adjusted with **Method 2** based on $V_m$ calculation, namely, $V_m=V_c - I_{Rs} * R_s$; **e)** eM (with $R_s$: 7.12 M$\Omega$) adjusted with **Method 1** based on actual $V_m$ measures within MatLab model; **f)** For comparison, the kinetics of the model were increased by a factor of 30, and stretched kinetics eliminated. eM (with $R_s$: 7.12 M$\Omega$) adjusted with **Method 1** based on actual $V_m$ measures. Fast kinetics provide a fast frequency response. Uncorrected eM is as in **a)**.